\documentclass[aps,prl,showpacs,twocolumn,groupedaddress]{revtex4}  

\usepackage{graphicx}  
\usepackage{dcolumn}   
\usepackage{bm}        
\usepackage{amssymb}   

\def\lsim{\mathrel{\rlap{\lower4pt\hbox{\hskip1pt$\sim$}}\raise1pt\hbox{$<$}}}
\def\gsim{\mathrel{\rlap{\lower4pt\hbox{\hskip1pt$\sim$}}\raise1pt\hbox{$>$}}}

\begin{document}

  
\title{Study of $Z\gamma$ events and limits on anomalous $ZZ\gamma$ and 
$Z\gamma\gamma$ couplings in $p\overline{p}$ collisions 
at $\sqrt{s} = 1.96$ TeV}
%
\author{                                                                      
V.M.~Abazov,$^{35}$                                                           
B.~Abbott,$^{72}$                                                             
M.~Abolins,$^{63}$                                                            
B.S.~Acharya,$^{29}$                                                          
M.~Adams,$^{50}$                                                              
T.~Adams,$^{48}$                                                              
M.~Agelou,$^{18}$                                                             
J.-L.~Agram,$^{19}$                                                           
S.H.~Ahn,$^{31}$                                                              
M.~Ahsan,$^{57}$
% ...
\\                                                                            
\vskip 0.30cm                                                                 
\centerline{(D\O\ Collaboration)}                                               
\vskip 0.30cm                                                                 
}                                                                             
\address{                                                                     
\centerline{$^{1}$Universidad de Buenos Aires, Buenos Aires, Argentina}       
\centerline{$^{2}$LAFEX, Centro Brasileiro de Pesquisas F{\'\i}sicas,         
                  Rio de Janeiro, Brazil}                                     
\centerline{$^{3}$Universidade do Estado do Rio de Janeiro,                   
                  Rio de Janeiro, Brazil}                                     
\centerline{$^{4}$Instituto de F\'{\i}sica Te\'orica, Universidade            
                  Estadual Paulista, S\~ao Paulo, Brazil}                     
\centerline{$^{5}$University of Alberta, Edmonton, Alberta, Canada,           
               Simon Fraser University, Burnaby, British Columbia, Canada,}   
\centerline{York University, Toronto, Ontario, Canada, and                    
         McGill University, Montreal, Quebec, Canada}                         
\centerline{$^{6}$Institute of High Energy Physics, Beijing,                  
                  People's Republic of China}                                 
\centerline{$^{7}$University of Science and Technology of China, Hefei,       
                  People's Republic of China}                                 
\centerline{$^{8}$Universidad de los Andes, Bogot\'{a}, Colombia}             
\centerline{$^{9}$Center for Particle Physics, Charles University,            
                  Prague, Czech Republic}                                     
\centerline{$^{10}$Czech Technical University, Prague, Czech Republic}        
\centerline{$^{11}$Institute of Physics, Academy of Sciences, Center          
                  for Particle Physics, Prague, Czech Republic}               
\centerline{$^{12}$Universidad San Francisco de Quito, Quito, Ecuador}        
\centerline{$^{13}$Laboratoire de Physique Corpusculaire, IN2P3-CNRS,         
                 Universit\'e Blaise Pascal, Clermont-Ferrand, France}        
\centerline{$^{14}$Laboratoire de Physique Subatomique et de Cosmologie,      
                  IN2P3-CNRS, Universite de Grenoble 1, Grenoble, France}     
\centerline{$^{15}$CPPM, IN2P3-CNRS, Universit\'e de la M\'editerran\'ee,     
                  Marseille, France}                                          
\centerline{$^{16}$Laboratoire de l'Acc\'el\'erateur Lin\'eaire,              
                  IN2P3-CNRS, Orsay, France}                                  
\centerline{$^{17}$LPNHE, IN2P3-CNRS, Universit\'es Paris VI and VII,         
                  Paris, France}                                              
\centerline{$^{18}$DAPNIA/Service de Physique des Particules, CEA, Saclay,    
                  France}                                                     
\centerline{$^{19}$IReS, IN2P3-CNRS, Universit\'e Louis Pasteur, Strasbourg,  
                France, and Universit\'e de Haute Alsace, Mulhouse, France}   
\centerline{$^{20}$Institut de Physique Nucl\'eaire de Lyon, IN2P3-CNRS,      
                   Universit\'e Claude Bernard, Villeurbanne, France}         
\centerline{$^{21}$III. Physikalisches Institut A, RWTH Aachen,               
                   Aachen, Germany}                                           
\centerline{$^{22}$Physikalisches Institut, Universit{\"a}t Bonn,             
                  Bonn, Germany}                                              
\centerline{$^{23}$Physikalisches Institut, Universit{\"a}t Freiburg,         
                  Freiburg, Germany}                                          
\centerline{$^{24}$Institut f{\"u}r Physik, Universit{\"a}t Mainz,            
                  Mainz, Germany}                                             
\centerline{$^{25}$Ludwig-Maximilians-Universit{\"a}t M{\"u}nchen,            
                   M{\"u}nchen, Germany}                                      
\centerline{$^{26}$Fachbereich Physik, University of Wuppertal,               
                   Wuppertal, Germany}                                        
\centerline{$^{27}$Panjab University, Chandigarh, India}                      
\centerline{$^{28}$Delhi University, Delhi, India}                            
\centerline{$^{29}$Tata Institute of Fundamental Research, Mumbai, India}     
\centerline{$^{30}$University College Dublin, Dublin, Ireland}                
\centerline{$^{31}$Korea Detector Laboratory, Korea University,               
                   Seoul, Korea}                                              
\centerline{$^{32}$CINVESTAV, Mexico City, Mexico}                            
\centerline{$^{33}$FOM-Institute NIKHEF and University of                     
                  Amsterdam/NIKHEF, Amsterdam, The Netherlands}               
\centerline{$^{34}$Radboud University Nijmegen/NIKHEF, Nijmegen, The          
                  Netherlands}                                                
\centerline{$^{35}$Joint Institute for Nuclear Research, Dubna, Russia}       
\centerline{$^{36}$Institute for Theoretical and Experimental Physics,        
                  Moscow, Russia}                                             
\centerline{$^{37}$Moscow State University, Moscow, Russia}                   
\centerline{$^{38}$Institute for High Energy Physics, Protvino, Russia}       
\centerline{$^{39}$Petersburg Nuclear Physics Institute,                      
                   St. Petersburg, Russia}                                    
\centerline{$^{40}$Lund University, Lund, Sweden, Royal Institute of          
                   Technology and Stockholm University, Stockholm,            
                   Sweden, and}                                               
\centerline{Uppsala University, Uppsala, Sweden}                              
\centerline{$^{41}$Lancaster University, Lancaster, United Kingdom}           
\centerline{$^{42}$Imperial College, London, United Kingdom}                  
\centerline{$^{43}$University of Manchester, Manchester, United Kingdom}      
\centerline{$^{44}$University of Arizona, Tucson, Arizona 85721, USA}         
\centerline{$^{45}$Lawrence Berkeley National Laboratory and University of    
                  California, Berkeley, California 94720, USA}                
\centerline{$^{46}$California State University, Fresno, California 93740, USA}
\centerline{$^{47}$University of California, Riverside, California 92521, USA}
\centerline{$^{48}$Florida State University, Tallahassee, Florida 32306, USA} 
\centerline{$^{49}$Fermi National Accelerator Laboratory, Batavia,            
                   Illinois 60510, USA}                                       
\centerline{$^{50}$University of Illinois at Chicago, Chicago,                
                   Illinois 60607, USA}                                       
\centerline{$^{51}$Northern Illinois University, DeKalb, Illinois 60115, USA} 
\centerline{$^{52}$Northwestern University, Evanston, Illinois 60208, USA}    
\centerline{$^{53}$Indiana University, Bloomington, Indiana 47405, USA}       
\centerline{$^{54}$University of Notre Dame, Notre Dame, Indiana 46556, USA}  
\centerline{$^{55}$Iowa State University, Ames, Iowa 50011, USA}              
\centerline{$^{56}$University of Kansas, Lawrence, Kansas 66045, USA}         
\centerline{$^{57}$Kansas State University, Manhattan, Kansas 66506, USA}     
\centerline{$^{58}$Louisiana Tech University, Ruston, Louisiana 71272, USA}   
\centerline{$^{59}$University of Maryland, College Park, Maryland 20742, USA} 
\centerline{$^{60}$Boston University, Boston, Massachusetts 02215, USA}       
\centerline{$^{61}$Northeastern University, Boston, Massachusetts 02115, USA} 
\centerline{$^{62}$University of Michigan, Ann Arbor, Michigan 48109, USA}    
\centerline{$^{63}$Michigan State University, East Lansing, Michigan 48824,   
                   USA}                                                       
\centerline{$^{64}$University of Mississippi, University, Mississippi 38677,  
                   USA}                                                       
\centerline{$^{65}$University of Nebraska, Lincoln, Nebraska 68588, USA}      
\centerline{$^{66}$Princeton University, Princeton, New Jersey 08544, USA}    
\centerline{$^{67}$Columbia University, New York, New York 10027, USA}        
\centerline{$^{68}$University of Rochester, Rochester, New York 14627, USA}   
\centerline{$^{69}$State University of New York, Stony Brook,                 
                   New York 11794, USA}                                       
\centerline{$^{70}$Brookhaven National Laboratory, Upton, New York 11973, USA}
\centerline{$^{71}$Langston University, Langston, Oklahoma 73050, USA}        
\centerline{$^{72}$University of Oklahoma, Norman, Oklahoma 73019, USA}       
\centerline{$^{73}$Brown University, Providence, Rhode Island 02912, USA}     
\centerline{$^{74}$University of Texas, Arlington, Texas 76019, USA}          
\centerline{$^{75}$Southern Methodist University, Dallas, Texas 75275, USA}   
\centerline{$^{76}$Rice University, Houston, Texas 77005, USA}                
\centerline{$^{77}$University of Virginia, Charlottesville, Virginia 22901,   
                   USA}                                                       
\centerline{$^{78}$University of Washington, Seattle, Washington 98195, USA}  
}                                                                             
\date{\today}

\begin{abstract}
We present a measurement of the $Z\gamma$ production cross section and limits 
on anomalous $ZZ\gamma$ and $Z\gamma\gamma$ couplings for form-factor 
scales of $\Lambda$ = $750$ and $1000$~GeV. The measurement is based on 
138 (152) candidate events in the $ee\gamma$ ($\mu\mu\gamma$) final state 
using 320 (290)~pb$^{-1}$ of $p\bar{p}$ collisions at $\sqrt{s} = 1.96$~TeV. 
The 95\%~C.L. limits on real and imaginary parts of individual anomalous 
couplings are $|h_{10,30}^{Z}|<0.23$, $|h_{20,40}^{Z}|<0.020$, 
$|h_{10,30}^{\gamma}|<0.23$, and  $|h_{20,40}^{\gamma}|<0.019$ 
for $\Lambda$ = 1000~GeV.  
\end{abstract}

\pacs{12.15.Ji, 13.40.Em, 13.85.Qk}
\maketitle

Studies of events containing pairs of vector bosons provide important tests of 
the standard model (SM) of electroweak interactions. In the SM, the trilinear 
gauge couplings of the $Z$ boson to the photon are zero; therefore, photons 
do not interact with $Z$ bosons at lowest order. Evidence for such an 
interaction would indicate new physics~\cite{higgsusy, Gounaris_2002za}. 

Studies of $Z$ boson and photon production have been made by the 
CDF~\cite{cdf} and D\O~\cite{d01} collaborations using $p\bar{p}$ 
collisions, and by the DELPHI~\cite{delphi}, L3~\cite{L3}, and 
OPAL~\cite{opal} collaborations using $e^+e^-$ collisions.
We present a new study of $Z \gamma$ production using $Z$ boson decays
to $e^{+}e^{-}$ and $\mu^{+}\mu^{-}$, where the dilepton system can be
produced by either an on-shell $Z$ boson, or a virtual $Z$ boson or $\gamma$
(the Drell-Yan process). The dilepton plus photon final state, 
$\ell^+\ell^-\gamma$, can be produced in the SM through either of 
two processes. The photon may be emitted through
initial state radiation (ISR) from one of the partons in the $p$ or 
$\bar{p}$, or produced as final state radiation (FSR)
from one of the final state leptons. We collectively refer to these 
processes as $Z\gamma$ production.

The SM $Z\gamma$ processes produce photons with a rapidly falling
transverse energy, $E_T^\gamma$. In contrast, anomalous 
$ZZ\gamma$ and $Z\gamma\gamma$ couplings, which
appear in extensions of the SM, can cause production 
of photons with high $E_T^\gamma$ and can increase the $\ell^+\ell^-\gamma$ 
cross section compared to the SM prediction. Below we describe a search 
for this anomalous production within the framework of Ref.~\cite{baur}.
This formalism assumes only that the $ZV\gamma$ ($V$=$Z$, $\gamma$) 
couplings are Lorentz- and gauge-invariant.
The most general $ZV\gamma$ coupling is parameterized by 
two CP-violating ($h_1^V$ and $h_2^V$) and two CP-conserving 
($h_3^V$ and $h_4^V$) complex coupling parameters. Partial wave 
unitarity is ensured at high energies by using a form-factor ansatz 
$h_{i}^{V} = h_{i0}^{V}/(1+\hat{s}/\Lambda^{2})^{n_i}$ ($i=1,...,$ 4), where 
$\sqrt{\hat{s}}$ is the parton center-of-mass energy, $\Lambda$ is the 
form-factor scale, and $n_{i}$ is the form factor power. 
We set the form factor powers $n_{1}=n_{3}=3$ and $n_{2}=n_{4}=4$, 
in accordance with~\cite{baur}.

The data are collected by the D\O~Run II detector at the Fermilab Tevatron 
Collider with $p\bar{p}$ center-of-mass energy $\sqrt{s}$ = 1.96~TeV 
between April 2002 and June 2004. The integrated luminosities used for 
this analysis are 320~pb$^{-1}$ for the electron final state 
and 290~pb$^{-1}$ for the muon final state.    

The D\O~detector~\cite{run2det} consists of an inner tracker, surrounded by 
liquid-argon/uranium calorimeters, and a muon spectrometer. 
The detector sub-systems provide measurements
over the full range of azimuthal angle $\phi$ and over different, 
overlapping regions of detector pseudorapidity $\eta$.
The inner tracker consists of a silicon microstrip tracker (SMT) 
and a central fiber tracker (CFT), both located within a 2~T 
superconducting solenoidal magnet. The CFT and the SMT have coverage out to
$|\eta| \lsim$~1.8 and $|\eta| \approx $~3.0, respectively.
The calorimeter is divided into a central calorimeter (CC) $|\eta| < $~1.1 
and two end calorimeters (EC) housed in separate cryostats which extend 
coverage to $|\eta|\approx$~4. The calorimeters are longitudinally segmented 
into electromagnetic (EM) and hadronic sections.
The muon system lies outside the calorimeters and consists of 
tracking detectors, scintillation trigger counters, and a 1.8~T toroid magnet.
It has coverage up to $|\eta| \approx $~2.0. Luminosity is measured 
using plastic scintillator arrays located in front of the EC cryostats, 
covering 2.7~$< |\eta| <$~4.4.

The data are collected with a three-level trigger system (L1, L2, and L3).
We require that the events in the electron decay channel
satisfy one of the high-$E_T$ single electron triggers, while the events in 
the muon decay channel must fire one of the high-$p_T$ single or 
dimuon triggers. The single electron triggers require a significant amount of 
energy deposited in the EM section of the calorimeter at 
L1. At L3, additional requirements are imposed on 
the fraction of energy deposited in the EM calorimeter 
and the shape of the energy deposition. 
The efficiency of the electron trigger requirement
is about 80\% for an electron with $E_T \approx$~25~GeV and 
more than 98\% for $E_T >$~30~GeV. The muon trigger requires 
hits in the muon system scintillator at L1, and in portions 
of the data set also requires spatially-matched hits in the muon 
tracking detectors. At L2, muon track segments are reconstructed 
and $p_{T}$ requirements are imposed. At L3, some of the triggers 
used in this analysis require muon candidate events to have a 
high-$p_T$ track reconstructed in the inner tracker. The logical OR 
of single and dimuon triggers has an efficiency of 92\% for muons 
from $Z$ boson decay.

Electrons are reconstructed as clusters of energy in the calorimeter.  
These clusters are required to have 90\% of their energy
deposited in the EM calorimeter (in either the central 
calorimeter $|\eta |< $~1.1, or the end calorimeter 1.5~$ < |\eta| < 2.5$).
We require that the longitudinal and transverse
shower shape of the cluster is consistent with that expected from an electron, 
and that the cluster is isolated from other activity in the 
calorimeter. Electron candidates in the central calorimeter
are required to have spatially matched tracks.
At least one electron candidate must be identified in the CC region 
and at least one is required to have $p_T >$~25~GeV/$c$. Muons are 
identified by a central track matched to segments in the muon 
system. The muon must be within $|\eta|<$~2.0. 
To reduce potential contamination from hadronic $b\overline{b}$ events, 
we impose isolation requirements on the
muon candidates in both the calorimeter and central tracker.
To remove the background from cosmic ray muons, muon tracks
must originate from the beam region and not be back-to-back.
$Z$ boson candidates are reconstructed by requiring a pair
of high-$p_T$ ($p_{T} > $~15~GeV/$c$) electrons or muons that form 
an invariant mass above 30~GeV/$c^2$.

In addition to a $Z$ boson candidate, we require events to have a photon 
candidate, with a separation from both of the leptons of 
$\Delta {\cal R}$~=~$\sqrt{(\phi_\ell - \phi_\gamma)^{2} + (\eta_{\ell}-\eta_{\gamma})^{2}}$~$>$~$0.7$ 
and with $E_T^\gamma >$~8~GeV. Photons are reconstructed as energy clusters
in the central calorimeter. The transverse shower shape of the cluster 
must be consistent with that expected from a photon. We also require a 
photon candidate to deposit at least 90\% of its energy in the EM 
calorimeter and to be isolated from other activity in the calorimeter 
and the tracker.

Muon and electron detection efficiencies for the above
requirements are determined using a sample of $Z\to\ell\ell$ events.
In the electron channel the combined trigger and reconstruction 
efficiency is measured to be (73~$\pm$~4)\%. In the muon channel 
it is measured to be (81~$\pm$~4)\%. The photon identification 
efficiency is measured as a 
function of $E_{T}^\gamma$ using a Monte Carlo simulation. A systematic
uncertainty of 4\% is assessed from the difference between the simulated
electrons and electron candidates in $Z\to ee$ data, and the difference
between simulated electrons and photons. 
The photon identification efficiency is $E_{T}$-dependent 
and rises from about 75\% at 8~GeV to about 90\% above 27~GeV.

Backgrounds from processes where the photon is real and one or both of the 
leptons are misidentified are found to be negligible. Contributions from 
$Z$($\rightarrow \tau^+\tau^-$)$\gamma$ events with leptonic decays 
of the tau are less than 1\% of the sample.
The only significant source of background to $Z\gamma$ production is 
from $Z$+jets processes in which a jet is misidentified as a photon. 
We estimate the $Z$+jets background by folding the jet-$E_T$ spectrum in
$Z$+jets events with the probability for a jet to be misidentified as a photon.
The probability is measured as a function of the photon candidate's $E_T$ 
using a sample of events dominated by QCD multijet processes. We correct
the misidentification probability for direct photon production ($\gamma$+jets)
by fitting the photon candidate $E_T$ distribution to the functional form 
derived in~\cite{RunIDirectGamma}. For low $E_T$ ($E_T <$~75~GeV) this 
contribution is measured to be 9\%, and we take this number as a systematic
uncertainty. The misidentification probability is about 
5 $\times 10^{-3}$ and decreases with $E_{T}$.  

\begin{figure}
\includegraphics[scale=0.45]{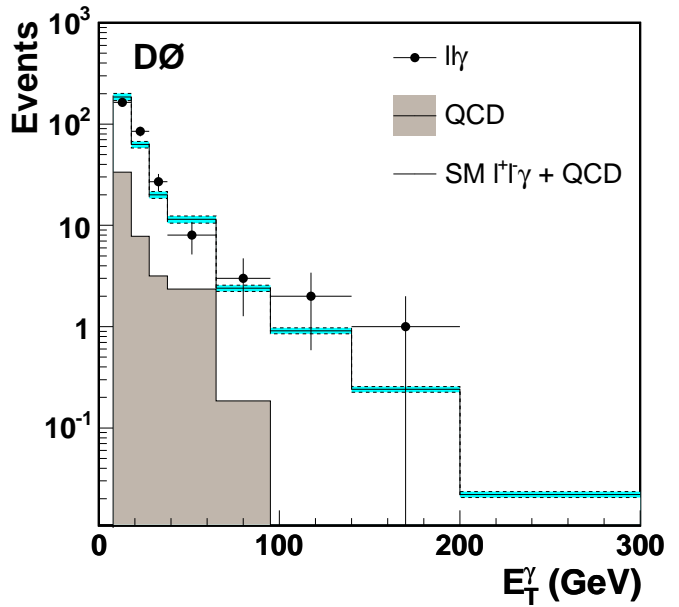}
\caption{
Photon candidate $E_{T}$ spectrum for $\ell\ell\gamma$ data (solid circles), 
QCD multijet background (shaded histogram), and the standard model 
plus background (histogram).
The shaded band is the systematic uncertainty on the SM plus background. 
The Monte Carlo distribution is normalized to the luminosity. 
\label{fig:ptg_all}}
\end{figure}
We use an event generator employing leading order (LO) QCD calculations
with a detector simulation tuned with $Z$ boson candidate events to
calculate the acceptances for the data and expected rates from both
the SM and anomalous $Z\gamma$ productions~\cite{baur}.
We use the CTEQ6L~\cite{cteq} parton distribution function (PDF) 
set. We estimate the uncertainty due to PDF choice using the prescription
in~\cite{cteqerror} to be 3.3\%.
Using a NLO $Z\gamma$ Monte Carlo~\cite{baurnlo} generator,
we calculate an $E_T^\gamma$-dependent $K$-factor
to parameterize the effect of $E_T^\gamma$-dependent NLO corrections in 
the LO Monte Carlo sample. The uncertainty due to the choice of 
$K$-factor (flat $vs.$ $E_T^\gamma$-dependent) is found to be
negligible.

We observe 138 events in the electron channel, to be compared to the SM 
estimate of $95.3 \pm 4.9$ $e^+e^-\gamma$ events and 
$23.6\pm 2.3$ background events. In the muon channel,
we observe 152 events $vs.$ an estimated $126.0 \pm 7.8$ SM $\mu^+\mu^-\gamma$
events and $22.4\pm 3.0$ background events. The uncertainty in the SM signal 
is dominated by the uncertainty in the lepton and photon reconstruction 
efficiencies, and that in the background estimation is dominated by the 
uncertainty in the jet misidentification probability.

\begin{figure}
\includegraphics[scale=0.45]{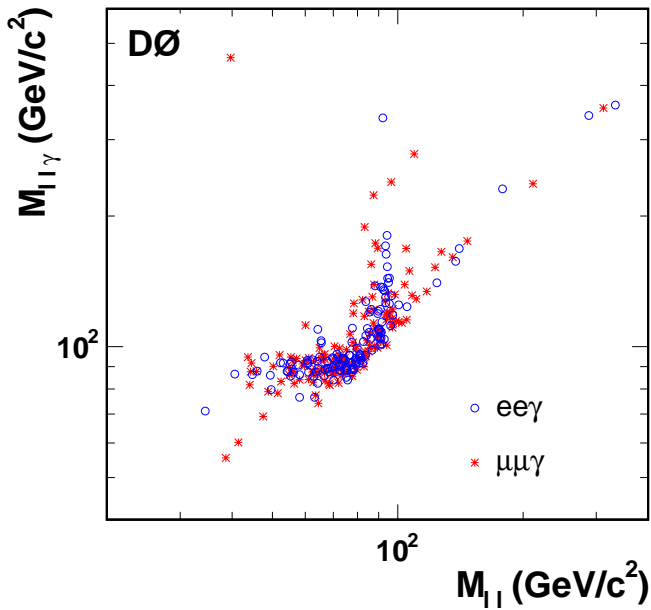}
\caption{Dilepton+photon $vs.$ dilepton mass of $Z\gamma$ candidate events.
Candidates in the electron channel are shown as empty circles,
while the muon mode candidates are shown as stars. 
\label{fig:ll_llg}}
\end{figure}

The $E_{T}$ spectrum for photon candidates is shown in Fig.~\ref{fig:ptg_all}
with the estimation of the total SM prediction and its QCD background
component overlaid. The highest transverse energy photon
in the electron channel is 105~GeV, while the highest transverse
energy photon in the muon channel is 166~GeV.
In Fig.~\ref{fig:ll_llg} we plot the three-body mass ($M_{\ell\ell\gamma}$)
against the dilepton mass ($M_{\ell\ell}$) for each event in the data. 
The ISR events with a dilepton system produced by an on-shell $Z$ boson 
populate a vertical band at $M_{\ell\ell}$ around $Z$ boson mass, $M_Z$,
and $M_{\ell\ell\gamma}>M_Z$. The on-shell $Z$ boson FSR events cluster
along a horizontal band at $M_{\ell\ell\gamma}=M_Z$ and have 
$M_{\ell\ell}<M_Z$. The Drell-Yan events populate the diagonal band 
with $M_{\ell\ell}\approx M_{\ell\ell\gamma}$ extending from the lower
left to the upper right corner of the plot.

For events satisfying $\Delta {\cal R}_{\ell\gamma} >$~0.7, 
$E_{T}^{\gamma} > $~8~GeV, and
$M_{\ell\ell} >$~30~GeV/$c^2$, the combined cross section is measured to be 
$4.2 \pm 0.4$~(stat+sys)$ \pm 0.3$~(lum)~pb, where 
the first uncertainty includes contributions from statistics and all
systematic effects except the luminosity, and the second
is due to the luminosity measurement uncertainty~\cite{d0lumi}. 
This value is in agreement with the expected value of 
$3.9 ^{+0.1}_{-0.2}$~pb from NLO theory calculations~\cite{baurnlo}.  

Given the separation exhibited in Fig.~\ref{fig:ll_llg},
we can measure a cross section of ISR-enhanced $Z\gamma$ production.
By minimizing the effect of final state radiation, any anomalous 
contributions from a trilinear boson vertex would become more apparent.
By requiring the dilepton mass and the three-body mass to 
exceed $65$~GeV/$c^2$ and $100$~GeV/$c^2$, respectively, 
(along with all previous requirements), the 
SM Monte Carlo simulation indicates that 80\% of the remaining events
are due to initial state radiation.
For this restricted sample we observe 55 and 62 events in the 
electron and muon channels, respectively. The cross section is 
measured to be 1.07~$\pm$~0.15(stat+sys)~$\pm$~0.07(lum)~pb, 
in agreement with the expected $0.94 ^{+0.02}_{-0.05}$~pb~\cite{baurnlo}.

\begin{figure}
\includegraphics[scale=0.45]{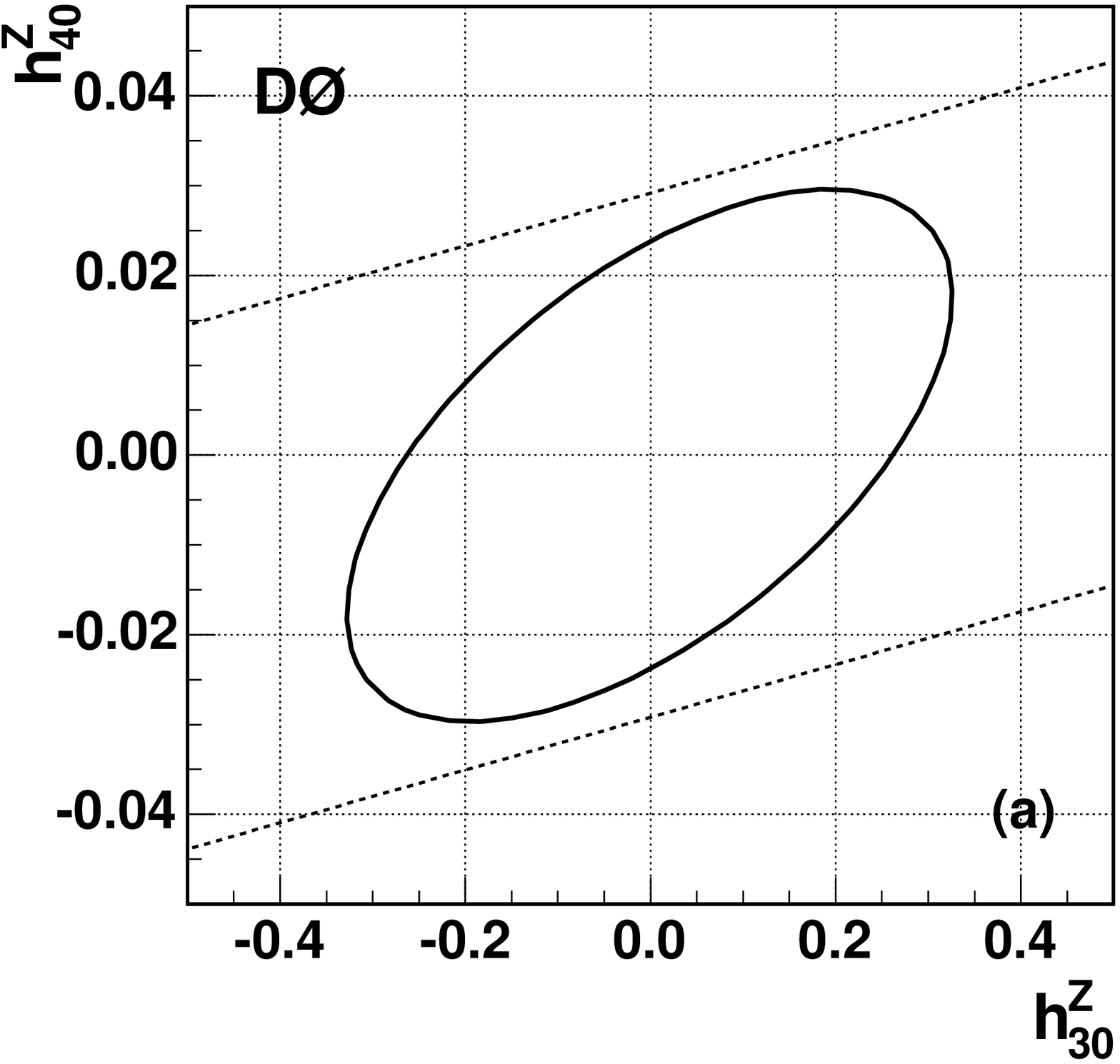}
\includegraphics[scale=0.45]{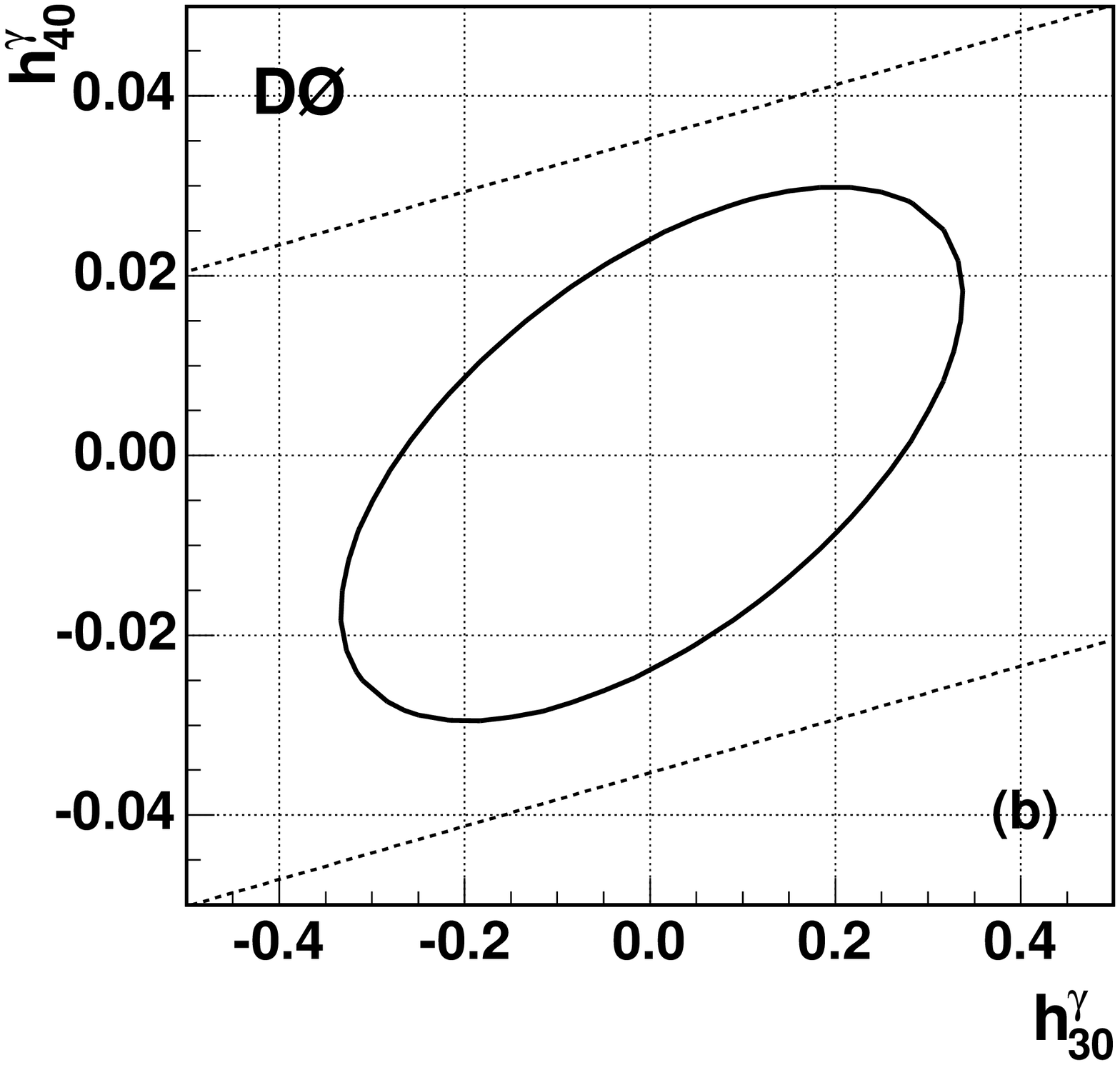}
\caption{The 95\% C.L. two-dimensional exclusion limits for CP-conserving
$ZZ\gamma$ (a) and $Z\gamma\gamma$ (b) couplings for $\Lambda$~=~1~TeV.
Dashed lines illustrate the unitarity constraints.
\label{fig:zgglimtev}}
\end{figure}

Given the good agreement observed between the data and the SM prediction,
we extract limits on anomalous couplings. We generate Monte Carlo events in 
a two-dimensional grid of CP-violating anomalous couplings ($h_{10}^{V}$ 
and $h_{20}^{V}$) and do the same for 
CP-conserving ($h_{30}^{V}$ and $h_{40}^{V}$) 
anomalous couplings. We calculate the likelihood of the agreement between 
the $E_T^\gamma$ distribution in data to the 
estimated background and Monte Carlo simulation for each point
of the grid. Assuming Poisson statistics for the data and Gaussian systematic 
uncertainties, we extract the 95\% C.L. limits on each of the 
anomalous couplings while assuming the others are zero. The limits on 
CP-violating and CP-conserving anomalous couplings are nearly identical. 
We also find the limits on real and imaginary parts of the couplings to be 
similar as well. We present the limits on both real and 
imaginary parts of the CP-conserving and CP-violating couplings in 
Table~\ref{tab:LimitSummary}. The two-dimensional limit contours
on individual CP-conserving couplings are shown in Fig.~\ref{fig:zgglimtev}.

\begin{table}
\caption{\label{tab:LimitSummary}Summary of the 95\% C.L. limits on
the anomalous couplings. Limits are set by allowing only 
the real or imaginary part of one coupling to vary; all others are fixed to 
their standard model values. As indicated, we find limits on CP-conserving and
CP-violating parameters to be nearly identical. We also find that nearly 
identical limits apply to the real or imaginary parts of all couplings.}
\begin{ruledtabular}
\begin{tabular}{ccc}
Coupling               & $\Lambda =$~750~GeV & $\Lambda =$~1~TeV\\ \hline
$|\Re {\it e} (h_{10, 30}^{Z})|$, $|\Im {\it m} (h_{10, 30}^{Z})|$          & 0.24                & 0.23 \\
$|\Re {\it e} (h_{20, 40}^{Z})|$, $|\Im {\it m}(h_{20, 40}^{Z})|$           & 0.027               & 0.020\\
$|\Re {\it e} (h_{10, 30}^{\gamma})|$, $|\Im {\it m}(h_{10, 30}^{\gamma})|$ & 0.29                & 0.23 \\
$|\Re {\it e} (h_{20, 40}^{\gamma})|$, $|\Im {\it m}(h_{20, 40}^{\gamma})|$ & 0.030               & 0.019\\
\end{tabular}
\end{ruledtabular}
\end{table}

In conclusion, we have studied a sample of 290 $\ell\ell\gamma$ events,
consistent with $Z\gamma$ production. This sample exceeds that previously 
collected by D\O~by an order of magnitude.
This is due to three times more integrated luminosity, an increased production
cross section associated with the 10\% higher center-of-mass energy, and
significant improvements in particle detection efficiency achieved
with the D\O~Run II upgrade. The $\ell\ell\gamma$ cross section is measured 
to be 4.2~$\pm$~0.4(stat+syst)~$\pm$~0.3(lum)~pb. After additional
selection requirements, most of the final state radiation is removed, leaving
the sample dominated by initial state radiation.
The cross section for this ISR-enhanced $Z\gamma$ production us measured to be
1.07~$\pm$~0.15(stat+syst)~$\pm$~0.07(lum)~pb.
These values are consistent with the SM expectations. We observe
no significant deviation from the SM expectation in the total cross section
or photon $E_T$ distribution, and thus extract limits on anomalous 
$Z\gamma$ couplings.  
The one dimensional limits at 95\%~C.L. for both CP-conserving and CP-violating
couplings (both real and imaginary parts) are $|h_{10, 30}^{Z}|<0.23$, 
$|h_{20, 40}^{Z}|<0.020$, $|h_{10, 30}^{\gamma}|<0.23$, 
and $|h_{20, 40}^{\gamma}|<0.019$ for $\Lambda$ = 1 TeV.  
These limits are substantially more restrictive than previous results which 
have been presented using this formalism~\cite{d01}. The limits on $h_{20}^V$
and $h_{40}^V$ are more than twice as restrictive as the combined
results of the four LEP experiments~\cite{lepewwg}.

%
We thank the staffs at Fermilab and collaborating institutions, 
and acknowledge support from the 
Department of Energy and National Science Foundation (USA),  
Commissariat  \` a l'Energie Atomique and 
CNRS/Institut National de Physique Nucl\'eaire et 
de Physique des Particules (France), 
Ministry of Education and Science, Agency for Atomic 
   Energy and RF President Grants Program (Russia),
CAPES, CNPq, FAPERJ, FAPESP and FUNDUNESP (Brazil),
Departments of Atomic Energy and Science and Technology (India),
Colciencias (Colombia),
CONACyT (Mexico),
KRF (Korea),
CONICET and UBACyT (Argentina),
The Foundation for Fundamental Research on Matter (The Netherlands),
PPARC (United Kingdom),
Ministry of Education (Czech Republic),
Canada Research Chairs Program, CFI,
Natural Sciences and Engineering Research Council and 
WestGrid Project (Canada),
BMBF and DFG (Germany),
Science Foundation Ireland,
A.P.~Sloan Foundation,
Research Corporation,
Texas Advanced Research Program,
Alexander von Humboldt Foundation,
and the Marie Curie Fellowships.
%

\end{document}